\documentclass[11pt]{amsart}

\usepackage{amssymb}
\usepackage{epsf}
\usepackage{a4wide}

\newcommand{\ts}{\hspace{0.5pt}}

\newcommand{\RR}{\mathbb{R}\ts}

\DeclareMathOperator{\sgn}{sgn}
\DeclareMathOperator{\Sgn}{Sgn}

\newcommand{\bs}[1]{\boldsymbol{#1}}

\begin{document}

\title[Trying to be rare]
{Freely forming groups:\ Trying to be rare}

\author{Michael Baake}
\address{Fakult\"at f\"ur Mathematik, Univ.\ Bielefeld, 
Box 100131, 33501 Bielefeld, Germany}
\email{mbaake@math.uni-bielefeld.de}
\urladdr{http://www.math.uni-bielefeld.de/baake/}

\author{Uwe Grimm}
\address{Dept.\ of Mathematics,
The Open University, Walton Hall, Milton Keynes MK7 6AA, UK}
\email{u.g.grimm@open.ac.uk}
\urladdr{http://mcs.open.ac.uk/ugg2/}

\author{Harald Jockusch}
\address{Fakult\"at f\"ur Biologie, Univ.\ Bielefeld, 
Box 100131, 33501 Bielefeld, Germany}
\email{h.jockusch@uni-bielefeld.de}
\urladdr{http://www.uni-bielefeld.de/biologie/Entwicklungsbiologie/personen.htm}


\begin{abstract} 
  A simple weakly frequency dependent model for the dynamics of a
  population with a finite number of types is proposed, based upon an
  advantage of being rare. In the infinite population limit, this
  model gives rise to a non-smooth dynamical system that reaches its
  globally stable equilibrium in finite time. This dynamical system is 
  sufficiently simple to permit an explicit solution, built piecewise 
  from solutions of the logistic equation in continuous time. It displays 
  an interesting tree-like structure of coalescing components.
\end{abstract}

\maketitle

\vspace*{-5mm}\thispagestyle{empty}
\section{Introduction}

Imagine a large, but finite population of $N$ individuals, with $N$
fixed.  Imagine further that they are divided into $M$ disjoint
groups, of type $i$ and size $n_{i}^{}>0$, $1\le i\le M$, so that
$n_{1}^{}+\ldots +n_{M}^{}=N$.  Let us assume that there is an
advantage in belonging to a {\em smaller}\/ group, but that changing
the group is only reasonable if individuals from different groups
meet. As a justification for this setting, we envision that each
individual might only know the size of its own group, but not
necessarily the size of any other one.  Consequently, a comparison
requires the exchange of information.  For simplicity, we assume
honesty in this process.

To specify the dynamics, we assume that, at a fixed rate (which we
may choose to be $1$ for simplicity), two
individuals chosen at random from the entire population meet and act
according to the following protocol. If both are from the same group,
or from groups of the same size, no action is taken; otherwise, the
individual from the larger group joins the smaller one, by changing
type accordingly. We assume that the advantage depends on group size
only, not on type. Note that a situation where $n_{j}^{}=n_{i}^{}+1$
is changed into $n_{i}^{}=n_{j}^{}+1$ is overall neutral, and also
neutral from the point of view of the acting individual.

In this short note, we consider the resulting dynamics in the
deterministic limit as $N\to\infty$, with a fixed number of types.
This approach is appropriate because the protocol is deterministic,
and no extinction can occur in the finite size situation. Therefore,
no stochastic effects of relevance are possible.  Consequently, we
shall look at the deterministic time evolution of the corresponding
(discrete) probability distribution $\{p_{i}^{}\mid 1\le i\le M\}$ of
the types.  This setting is known as the \emph{infinite population
limit}\/ (IPL), compare \cite[Ch.\ 11.2]{EK} for background material
and \cite{BB,RB,GG} for recent examples. For a finite population,
$p_{i}^{}=n_{i}^{}/N$ are the type frequencies. The model thus shows a
frequency dependence, which is weak in the sense that only the signs
of the differences $p^{}_{i} - p^{}_{j}$ matter, but not their
magnitudes.

This approach leads to a simple, but rather interesting system of
nonlinear ODEs, with discontinuous right-hand side. The latter
property requires some care concerning the concepts of a solution and
of an equilibrium, since the standard results of ODE theory do not
directly apply.  However, this will cause no problem in our case,
where we shall see that the globally stable equilibrium,
$p^{}_{1}=\ldots=p^{}_{M}=1/M$, is reached in {\em finite}\/ time
from any initial condition in the interior of the probability
simplex (see below for details).

Our simple example is illustrative in the sense that a biologically
relevant, non-smooth dynamical system can be looked at and solved
explicitly, without invoking the general theory \cite{F} or resorting
to abstract results \cite{D}. In fact, to derive the solution,
one only needs a good understanding of the logistic equation in
continuous time, compare \cite[Ch.~1]{A}. In our multi-dimensional
system, the solution curves of the various components display a
tree-like coalescing structure. The corresponding coalescing times are
the relevant time scales that would also show up in smooth
approximations of this non-smooth dynamical system (where they appear
as the characteristic times of the exponential closing-in of the
components).

The model is a toy example of game theoretic flavour, compare
\cite{HS}, where a key ingredient is the use of non-local information
(the group sizes). As such, it can be considered as a model of
population dynamics \cite{HS,CKG}, but is rather different from the
standard models of population genetics, compare \cite{B,E}. As
possible applications, one could think of population patches (spatial
advantage), food specialisation (supply advantage), advertising
strategies (promotional advantage) or name choosing (recognisability),
to name but a few.

\smallskip
Similar models, and also more general ones, have been investigated in
mathematical game theory. One early example is the ideal free
distribution of Fretwell and Lucas in ecology \cite{FL}, which
describes the spatial distribution of a population among a fixed
number of resource patches. This is related to the evolutionary stable
strategies of game theory, see \cite{CKG} for a detailed discussion.
Various dynamics, including imitation and replicator dynamics, are
considered in this context, compare the excellent review by Hofbauer
and Sigmund \cite{HS2} with its extensive bibliography.

Even though our model can be seen as a special case of imitation
dynamics, compare \cite[Sec.~8.1]{HS} and \cite[Sec.~3.2]{HS2},
it results from a specific finite size model via the infinite 
population limit, and is thus fixed. Since we see this as a first 
step towards more realistic models including stochastic and
finite-size effects, we find it slightly more natural to use 
the classic notions of dynamical systems rather than to formulate 
the model and its solution in the language of game theory (we
shall give the translation below for comparison).

\bigskip
\section{The dynamical systems and their equilibria}

The reservoir of types, or {\em state space}, is $\{1, \ldots , M\}$,
i.e., there are {\em finitely}\/ many different states.  In the IPL
limit \cite{EK,BB,RB}, one considers the deterministic time evolution
of probability vectors $\bs{p}=(p^{}_1, \ldots , p^{}_M)$ with
$p^{}_i \ge 0$ and $\sum_i p_i = 1$, where $p_i$ is the proportion of
the (infinite) population in state $i$.  For the model described
above, with $\bs{p}=\bs{p}(t)$, this gives the system of ODEs
\begin{equation} \label{model1}
   \dot{p}_m \; = \; \sum_n p_m p_n \, \sgn(p_n - p_m)
\end{equation}
where $\sgn$ is the signum function, i.e.,
\[
   \sgn(x) \; = \; \begin{cases} \hphantom{-}1, & x>0 \\
             \hphantom{-}0, & x=0 \\ -1, & x<0\, . \end{cases}
\]
More generally, one could consider the ODEs
\begin{equation} \label{model2}
   \dot{p}_m \; = \; \sum_n p_m p_n \, \varphi(p_n - p_m)
\end{equation}
where $\varphi \! : \; [-1,1] \longrightarrow \RR$ is
an odd function (i.e., $\varphi(-x)=-\varphi(x)$) that
is continuous, or (as $\sgn$ above) piecewise continuous
with only finitely many jumps. In particular, $\varphi(0)=0$.

In general, such discontinuities require solutions to be defined
accordingly, e.g., in the sense of Filippov \cite[\S 4]{F}, thus
referring to the well developed theory of ODEs with discontinuous
right-hand sides. Fortunately, and not untypically for models with a
biological context, one can introduce all ingredients required for our
model in an elementary way. Moreover, one can make most results
explicit, thus illustrating aspects usually obtained by qualitative
theory.

Let us first assume that $\varphi$ is continuous. Then,
the forward flow of the dynamical system \eqref{model2} leaves 
the cone of positive vectors invariant, which follows from
the observation that $p_m (t)=0$ implies $\dot{p}_m (t)=0$,
together with standard arguments from \cite[Ch.\ 16]{A}.
Since $\varphi(0)=0$, one also finds
\[   \sum_{m,n} p_m p_n \, \varphi(p_n - p_m) \; = \;
     \sum_{m < n} p_m p_n \, 
     \big( \varphi(p_n - p_m) + \varphi(p_m - p_n) \big)
     \; = \; 0,
\]
whence $\frac{\rm d}{{\rm d}t} (p^{}_1 + \ldots + p^{}_M) =0$ and
total mass is preserved. In particular, the forward flow of
\eqref{model2} preserves the simplex of probability vectors,
$\mathcal{P}_M := \{ \bs{p}\in\RR^M \mid p_i \ge 0 
\mbox{ and } \sum_i p_i =1 \}$. 
This is in line with the probabilistic interpretation
chosen above. 

In an appropriate generalisation, this remains true for the dynamical
system \eqref{model1} with its discontinuous right-hand side.  In
general, this claim requires some care, and is usually substantiated
by viewing the ODE \eqref{model1} within the scheme of so-called {\em
  differential inclusions}, where one considers equations of the form
$\dot{p}^{}_n \in F(t,\bs{p})$ with set-valued right-hand sides
\cite[Sec.\ 2]{K}, see also \cite[Sec.~3.3]{HS2} for its appearance in
game theory, e.g., in the context of the so-called best response
dynamics.  {}For our system \eqref{model1}, the function $\sgn(x)$
would simply be replaced by the set-valued function $\Sgn$, defined by
$\Sgn(x) = \{\sgn(x)\}$ for $x\ne 0$ together with $\Sgn(0) = [-1,1]$,
compare \cite[Ex.\ 2.1.1]{K}.  Since our system is autonomous and
well-behaved, it is then essentially straight-forward, following
\cite[Thms.\ 2.2.1 and 2.2.2]{K}, to establish the existence and
uniqueness result for its solution in forward time. Fortunately,
Eq.~\eqref{model1} permits a much simpler approach, based directly on
the single-valued function $\sgn(x)$.

Clearly, solutions of \eqref{model1} can no longer be expected to be
differentiable everywhere. The starting point is then the class of
absolutely continuous functions, which are differentiable almost
everywhere. A solution of our ODE \eqref{model1} is then a function of
this class that satisfies \eqref{model1} almost everywhere.  As we
shall see explicitly below, the solutions of \eqref{model1} are
actually piecewise smooth, with only {\em finitely\/} many points of
non-differentiability (coinciding with the coalescence points of
components). More precisely, at the glueing points, the value of the
previous segment provides the initial value for the next. It will
always be evident that this construction yields the unique solution.

In what follows, we shall only consider the dynamical system
restricted to the probability simplex $\mathcal{P}_M$. Moreover, we mainly
consider initial values from its interior, because otherwise, in view
of the right-hand side of \eqref{model1}, we could start with a
smaller state space right away, i.e., with a probability simplex
of smaller dimension. In this sense, the restriction to the interior
of $\mathcal{P}_M$ covers all cases, by suitable choice of $M$.

The equilibria of our systems \eqref{model1} or \eqref{model2} certainly
include the uniform distribution on $\{1, \ldots , M\}$ or that on any
subset of it. These correspond to the centre of the simplex $\mathcal{P}_M$ 
or to the centre of some boundary of it, the latter effectively being
a simplex $\mathcal{P}_{M'}$ with $M' < M$.
Further equilibria can exist if $\varphi (x)$ has zeros other than
$x=0$. However, we are primarily interested in stable equilibria, such
as $p_i \equiv 1/M$ for \eqref{model1}. In fact, the latter is even
\emph{globally stable}, meaning that any initial condition from the
\emph{interior}\/ of $\mathcal{P}_M$ will converge, under the forward
flow, to this equilibrium, compare \cite[Sec.~2.2]{HS2} for a brief
discussion of these concepts.

Let us briefly expand on this point. Observing that both models
\eqref{model1} and \eqref{model2} can be viewed as special cases 
of imitation dynamics in game theory (with the negative identity
matrix being the payoff matrix, compare \cite[Eq.~8.3]{HS}), one
can derive that they possess a Lyapunov function, subject to certain
conditions on the function $\varphi$. This can also be seen
explicitly as follows. If we define $f(t) := \sum_m \big(p_m (t)\big)^2$,
a simple double summation argument using \eqref{model2} shows that
\begin{equation} \label{lapu}
  \dot{f} \; = \; - \sum_{m,n} p_m p_n \,
  (p_n - p_m) \, \varphi (p_n - p_m) \, .
\end{equation}
Consequently, as all $p_m\ge 0$, $\dot{f} (t)\le 0$ if $\varphi$ is a
non-decreasing odd function (note that $x\,\varphi(x)$ is then even
and $\ge 0$).

If $x=0$ is the only zero of $\varphi(x)$ and $\bs{p} (t)$ a point in
the interior of $\mathcal{P}_M$, one has $\dot{f} (t) < 0$ unless
$\bs{p} (t)$ is the centre of $\mathcal{P}_M$. This establishes $f$
as a strict Lyapunov function for the interior of $\mathcal{P}_M$ in 
these cases, see \cite{A} for details on this concept. This observation
establishes the global stability result of the equilibrium point
$p^{}_1 = \ldots = p^{}_M = 1/M$, both for \eqref{model1} and
\eqref{model2}. The hierarchy of boundaries of $\mathcal{P}_M$ can
be treated separately, by identifying them with appropriate cases
of smaller dimension.

\bigskip
\section{Solution of the case $M=2$}

Let us consider Eq.~\eqref{model1} for $M=2$. Setting $p^{}_1 = p$ and
$p^{}_2 = q = 1-p$, it is clear that it now suffices to consider the
single ODE
\begin{equation}  \label{model3}
   \dot{p} \; = \; p (1-p) \, \sgn(1-2p) \, .
\end{equation}
If $p(t) < 1/2$, this simplifies to $\dot{p} = p (1-p)$, which is the
well-known logistic equation in continuous time, compare \cite[Ch.~1]{A}. 
It has the solution
\[ 
    p(t) = \frac{1}{1 + \alpha e^{(t_0 - t)}}
    \qquad \mbox{with} \quad \alpha = \frac{1-p(t_0)}{p(t_0)}\, . 
\]
If the initial condition of \eqref{model3} is $0 < p(t_0) < 1/2$, this
formula also provides the solution we need, until $p(t)$ reaches the
value $1/2$, whereupon the solution continues as $p(t)\equiv 1/2$ ---
the globally stable equilibrium.  As is not untypical of ODEs with
discontinuous right-hand sides, compare \cite[\S 15]{F}, this happens
after a {\em finite}\/ time, namely at
\[
    t_1 \; = \; t_0 + \log \frac{1-p(t_0)}{p(t_0)} 
        \; = \; t_0 + \log (\alpha) .
\]

Similarly, if $p(t) > 1/2$, Eq.~\eqref{model3} simplifies to
$\dot{p} = - p (1-p)$. Provided that $1 > p(t_0) > 1/2$, the 
solution now is
\[
   p(t) = \frac{1}{1 + \alpha e^{(t - t_0)}},
\]
with $\alpha$ as before, until $p(t)$ hits $1/2$ from above. This
happens at time
\[
    t_2 \; = \; t_0 + \log \frac{p(t_0)}{1-p(t_0)}
        \; = \; t_0 - \log (\alpha) .
\]
In this rather simple case, $q=1-p$, and the complete solution is thus
given in closed form, with one coalescence point at time $t_{1}$ or
$t_{2}$, depending on the initial condition. If $\alpha=1$ (i.e.,
$p(t_{0})=1/2$), $t_{1}=t_{2}=t_{0}$, and one is in the equilibrium
from the very beginning.

\bigskip
\section{Some comments on the general structure}

If $M>2$, one can simplify the situation enormously by assuming,
without loss of generality, that the initial condition is
non-degenerate (meaning that all types occur) and ordered, so that
\begin{equation} \label{order1}
   0 < p^{}_1 \le p^{}_2 \le \ldots \le p^{}_M < 1 \, .
\end{equation}
If the $p_i$ are all equal, we must have $p_i\equiv 1/M$, i.e., we are
(and stay) in equilibrium. So, let us assume that at least one further
inequality sign is proper, and that the leftmost position where
this happens is between $\ell$ and $\ell + 1$. In other words, 
\[
   0 < p^{}_1 = \ldots = p^{}_{\ell} < p^{}_{\ell + 1}
   \le \ldots \le p^{}_M < 1 \, .
\]
Since $\sum_{j}p_{j}=1$, one then has $\sum_{j=\ell+1}^{M}p_{j}=1-\ell
p_{i}$, for any $1\le i\le\ell$. Then, a short calculation with
Eq.~\eqref{model1} shows that the time evolution of all $p_i$ with
$1\le i\le \ell$ is identical, and given by the ODE
\begin{equation} \label{model4}
   \dot{p} \; = \; p (1 - \ell p)\, ,
\end{equation}
as long as $p^{}_{\ell} < p^{}_{\ell + 1}$ remains true.
This variant of the logistic equation has the solution
\[ 
   p(t) \; = \; \frac{1}{\ell + \beta e^{(t_0 - t)}}
   \qquad \mbox{with} \quad \beta = \frac{1}{p(t_0)} - \ell \, .
\]

In general, the situation can be described as follows.  Whatever the
initial condition is, one can bring it to the standard form
\eqref{order1}, by a permutation of the types. In forward time, the
ordering then remains that way. Step by step, two or more neighbouring
$p_i$ coalesce, all in finite time, until the equilibrium is reached
(also in finite time). With increasing $M$, there is a growing number
of different possibilities for the emerging coalescent trees of
components (not to be mixed up with the result of the well-known
coalescent process from population genetics, compare \cite{B,E}),
which makes a closed formula for the solution cumbersome. The general
picture, however, is rather intuitive, and will become evident from
the explicit solution for $M=3$.

\bigskip
\section{Solution of the case $M=3$}

\begin{figure}
\centerline{\epsfbox{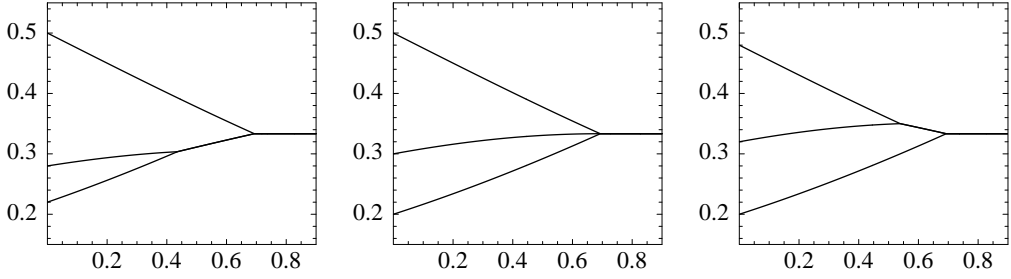}}
\caption{Examples of the non-degenerate classes of solutions for $M=3$. 
  The solutions $\big(p^{}_1 (t),p^{}_2 (t),p^{}_3 (t)\big)$ are drawn
  on top of each other, against time on the horizontal axis.
  Initial conditions at $t_{0}=0$
  are, from left to right, $(0.22,0.28,0.5)$, $(0.2,0.3,0.5)$ and
  $(0.2,0.32,0.48)$. The equilibrium
  $(\tfrac{1}{3},\tfrac{1}{3},\tfrac{1}{3})$ is reached at $t=\log(2)$
  in all three cases. In the left (right) frame, the first coalescence
  happens at $t=\log(\tfrac{17}{11})$ (at
  $t=\log(\tfrac{12}{7})$).
\label{dreipic}}
\end{figure}

Let us consider our system \eqref{model1} for $M=3$ and assume
\[  0 < p^{}_1 < p^{}_2 < p^{}_3 < 1  \]
in order to exclude any degenerate situation.
It is clear that we now have three different coalescent trees
to distinguish, see Figure \ref{dreipic}.

The system of ODEs now reads
\begin{equation} \label{model5}
   \dot{p}^{}_1  =  p^{}_1 (1 - p^{}_1) \; , \quad
   \dot{p}^{}_2  =  p^{}_2 (p^{}_3 - p^{}_1) \; , \quad 
   \dot{p}^{}_3  =  - p^{}_3 (1 - p^{}_3) \, .
\end{equation}
With $\alpha_i = (1-p_i(t_0))/p_i(t_0)$, the solutions 
$p^{}_1$ and $p^{}_3$ start as
\begin{equation} \label{sol13}
    p^{}_1 (t) = \frac{1}{1 + \alpha^{}_1 e^{(t_0-t)}} \; , \quad
    p^{}_3 (t) = \frac{1}{1 + \alpha^{}_3 e^{(t-t_0)}}
\end{equation}
while the solution for $p^{}_2$ can be obtained from here by
insertion and elementary integration,
\begin{equation} \label{sol2}
    p^{}_2 (t) = p^{}_2 (t_0)
    \frac{(1+\alpha^{}_1)(1+\alpha^{}_3)}
    {(1+\alpha^{}_1 e^{(t_0-t)})(1+\alpha^{}_3 e^{(t-t_0)})}\, .
\end{equation} 
These solutions remain valid until $p^{}_1$ meets with $p^{}_2$
or $p^{}_2$ with $p^{}_3$ (whatever is first), or both (if they
coincide). A simple calculation gives the coalescing times
\[
    t_{12} = t_0 + \log 
    \left( 1 + \frac{p^{}_2 - p^{}_1}
         {p^{}_1 ( p^{}_1 + p^{}_2)}\right)
    \quad \mbox{and} \quad
    t_{23} = t_0 - \log 
    \left( 1 - \frac{p^{}_3 - p^{}_2}
         {p^{}_3 ( p^{}_2 + p^{}_3)}\right) ,
\]
where the $p_i$ are to be read as $p_i (t_0)$.

If $t_{12}=t_{23}$, all three curves $p_i (t)$ coalesce in one point,
$(\tfrac{1}{3},\tfrac{1}{3},\tfrac{1}{3})$. From here, the solution is
then $p_i (t) \equiv \tfrac{1}{3}$, the globally stable equilibrium of
this situation, see the middle example in Figure~\ref{dreipic}.  This
happens if and only if the initial conditions satisfy
\begin{equation} \label{join}
    p^{}_2 = \frac{3 p^{}_1 (1-p^{}_1)}
             {1 + 3 p^{}_1} 
\end{equation}
in addition to $p^{}_1+p^{}_2+p^{}_3=1$.

Otherwise, we have to distinguish the cases $t_{12}<t_{23}$ and
$t_{12}>t_{23}$ (which correspond to replacing equality in
Eq.~\eqref{join} by $<$ or $>$, respectively). In the former case, the
above solutions for $p^{}_1(t)$ and $p^{}_2(t)$ are valid until
$t=t_{12}$. Afterwards, $p^{}_{12}(t)=p^{}_1(t)=p^{}_2(t)$ satisfies
Eq.~\eqref{model4} with $\ell=2$, and solution
\[
    p^{}_{12}(t) \; =\;    \frac{1}{2 + \beta e^{(t_{12}-t)}} \, ,
\quad\mbox{with}\quad \beta=\frac{1}{p^{}_{1}(t_{12})}-2 \, . 
\]
The solution for $p^{}_3(t)$ from \eqref{sol13} remains valid until it meets
with $p^{}_{12}(t)$, which happens at
\[
    t_{12|3} \; = \; t_{0} + 
    \log\frac{2 p^{}_{3}(t_{0})}{1-p^{}_{3}(t_{0})}\, .
\]
This time can easily be calculated from the condition
$p^{}_3(t_{12|3})=\tfrac{1}{3}$.

In the remaining case, $t_{12}>t_{23}$, we have 
\[
    p^{}_{23}(t) \; =\;   \frac{1}{2 + \beta e^{(t-t_{23})}} \, ,
\quad\mbox{with}\quad \beta=\frac{1}{p^{}_{3}(t_{23})}-2 \, ,
\]
after $p^{}_2(t)$ and $p^{}_3(t)$ met at $t=t_{23}$, until
$p^{}_{23}(t)$ meets with $p^{}_1(t)$ given in Eq.~\eqref{sol13} at
time
\[
    t_{1|23} \; = \; t_{0} + 
    \log\frac{1-p^{}_{1}(t_{0})}{2 p^{}_{1}(t_{0})} \, ,
\]
obtained from the condition $p^{}_1(t_{1|23})=\tfrac{1}{3}$.
Figure~\ref{dreipic} shows a representative example, from which the
general situation is clear.

\bigskip
\section{Concluding Remarks}

Equilibria of smooth dynamical systems are usually reached only
asymptotically. The above toy model does so in finite time, which is a
consequence of the discontinuous right-hand side of \eqref{model1}. If
the latter were replaced by a smooth approximation (e.g., a sigmoid
curve), neighbouring lines would no longer meet in finite time, but
close in exponentially fast instead -- otherwise, the qualitative
picture remains the same. In particular, if \eqref{model2} is considered
with an odd function $\varphi$ that is strictly monotonically increasing 
(hence including sigmoid curves), the nature of the globally stable
equilibrium remains unchanged. An advantage of the discontinuous system is
its exact solvability, including the calculation of all relevant time
scales.

As to the relevance of our findings for large, but {\em finite}\/
systems, fluctuations can lead to violations of the order relation
\eqref{order1}. However, it is clear that the convergence to the
equilibrium qualitatively remains the same if $N$ is a multiple of
$M$. Otherwise, the equilibrium is reached up to fluctuations of at
most one individual per group, i.e., of order $1/N$.  This mirrors the
deviations from the expectation values (over all realisations of the
stochastic process), the latter coinciding with the equilibrium of the
IPL limit in this simple case.

It might be interesting to extend the above model by means of a
stochastic (rather than deterministic) protocol, and to compare it
with related models of game theory \cite{HS,CKG}. This will
automatically lead to various extension and generalisations. To reach
a more complete and realistic picture, it will be necessary to develop
a proper stochastic frame and to investigate relevant finite size
effects by suitable simulations, see also \cite{HS2} for a brief
discussion of these issues.  These two aspects will certainly require
more attention in the future.

\bigskip
\bigskip
\section*{Acknowledgements}

It is our pleasure to thank E.~Baake and J.~Hofbauer for helpful 
criticism and W.-J.~Beyn for bringing reference \cite{K} to our 
attention. Financial support from German
Academic Exchange Service (DAAD, to M.B.), British Council (project
ARC 1213, to U.G.)  and Fonds der Chemischen Industrie (FCI, to H.J.)
is gratefully acknowledged.

\end{document}